\documentclass[prx,aps,showpacs,twocolumn,preprintnumbers,
amsmath,amssymb,superscriptaddress,longbibliography,colorlinks=true, a4paper=true, pdfstartview=FitV,linkcolor=blue, citecolor=blue, urlcolor=blue]{revtex4-1}

\usepackage[T1]{fontenc}		
\usepackage[english]{babel}		
\usepackage[utf8]{inputenc}	
\usepackage{times}
\usepackage{tikz}
\usepackage{tensor}
\usetikzlibrary{knots}

\usepackage{amsmath}			
\usepackage{amssymb}			
\usepackage{bbm}				

\usepackage{hyperref}
\usepackage{pbox}

\usepackage{graphicx}			
\usepackage{graphics}			
\DeclareGraphicsExtensions{.pdf}
\usepackage{color}	
\usepackage{makecell}

\newcommand{\bea}{\begin{eqnarray}}
\newcommand{\eea}{\end{eqnarray}}

\newcommand{\bk}{\mathbf{k}}

\renewcommand{\i}{\mathrm{i}}

\DeclareMathOperator{\HH}{\mathcal{H}}
\DeclareMathOperator{\Z}{\mathbb{Z}}

\newcommand{\PT}{\mathcal{PT}}

\newcommand{\tr}{\mathop{\mathrm{tr}}}
\newcommand{\bolds}[1]{\boldsymbol #1}

\allowdisplaybreaks

\begin{document}

\title{
Symmetry-protected exceptional and nodal points in non-Hermitian systems}

\author{Sharareh Sayyad}
\email{sharareh.sayyad@mpl.mpg.de}
\affiliation{Max Planck Institute for the Science of Light, Staudtstra\ss e 2, 91058 Erlangen, Germany}
\author{Marcus St{\aa}lhammar}
\affiliation{Department of Physics, Stockholm University, AlbaNova University Center, 106 91 Stockholm, Sweden}
\author{Lukas R\o dland}
\affiliation{Department of Physics, Stockholm University, AlbaNova University Center, 106 91 Stockholm, Sweden}
\author{Flore K. Kunst}
\affiliation{Max Planck Institute for the Science of Light, Staudtstra\ss e 2, 91058 Erlangen, Germany}

\begin{abstract}

One of the unique features of non-Hermitian~(NH) systems is the appearance of non-Hermitian degeneracies known as exceptional points~(EPs). The extensively studied defective EPs occur when the Hamiltonian becomes non-diagonalizable. Aside from this degeneracy, we show that NH systems may host two further types of degeneracies, namely, non-defective EPs and ordinary (Hermitian) nodal points. The non-defective EPs manifest themselves by i) the diagonalizability of the NH Hamiltonian at these points, ii) the non-diagonalizability of the Hamiltonian along certain intersections of these points and iii) instabilities in the Jordan decomposition when approaching the points from certain directions.
We demonstrate that certain discrete symmetries, namely parity-time, parity-particle-hole, and pseudo-Hermitian symmetry, guarantee the occurrence of both defective and non-defective EPs. We extend this list of symmetries by including the NH time-reversal symmetry in two-band systems. Two-band and four-band models exemplify our findings. Through an example, we further reveal that ordinary nodal points may coexist with defective EPs in NH models when the above symmetries are relaxed.
\end{abstract}

\maketitle

\paragraph*{Introduction.}Despite violating the axioms of quantum mechanics, non-Hermitian~(NH) Hamiltonians 
offer compelling descriptions for numerous interacting/open systems in various fields of physics~\cite{Bender2005,NHbook,Alexandre2020,Sayyad2021a,Sayyad2021}. The underlying physics of these effective Hamiltonians goes beyond the realm of Hermitian physics and has been immensely studied lately~\cite{NHreview, NHarc, BerryDeg, Heiss,EPrings, koziifu, CaBe2018, EPringExp, ourknots,ourknots2,chinghuaknot,yangknot, Ghorashi2021, Ghorashi2021b}. Aside from unraveling rich physics, the properties of NH systems are well-reflected in abstract mathematical frameworks, including homotopy theory~\cite{WojcikHomotopy, HuKnots, LiHomotopical, WojcikEigenvalue, HuKnot} and K-theory~\cite{GongTopological, ZhouPeriodic, KawabataSymmetry}. These frameworks provide a reliable toolbox to understand the exotic properties of NH systems and distinguish their behavior from Hermitian physics.

Noticeable distinctions of NH Hamiltonians from their Hermitian counterparts include the appearance of exceptional points~(EPs), as NH degeneracies. While one generally should satisfy $n-1$ complex constraints to realize EPs of order $n$ (EP$n$s), at which the Hamiltonian casts a Jordan block, recent studies have shown that the presence of certain discrete symmetries, such as parity-time~($\PT$), parity-particle-hole~($\cal CP$), or pseudo-Hermiticity~($\rm psH$) symmetry, reduces the total number of constraints~\cite{Budich2019, Yoshida2019, Okugawa, Zhou, HCphot, Etorus, 3Tsuneya, ips3, 4EPMarcus, Crippa21, Sayyad2022}.
Although the main focus of these studies was on characterizing symmetry-induced restrictions on defective EPs, drawing a link between discrete symmetries and non-defective degeneracies associated with the NH Hamiltonian beyond case studies have received little attention. 
Furthermore, while there is a near consensus on calling defective degeneracies~\cite{Xu2017, Kawabata2019} defective EPs~\footnote{ Except in Ref.~\cite{Shen2018} where `hybrid points' have also been introduced based on the asymptotic dispersion relations close to defective degeneracies. Note that branch cuts do not terminate at these points~\cite{Yang2021}.}, non-defective degeneracies in NH systems are referred to as diabolic points~\cite{Keck2003, Xue2020, Wiersig2022, Cui2022}, Fermi points~\cite{Yang2021}, Dirac points and vortex points~\cite{Shen2018}. These non-defective degeneracies are different in nature, as both Dirac points and vertex points might appear in the vicinity of defective EPs, while diabolic points and Fermi points are similar to (Hermitian) ordinary nodal points (ONPs).

Recalling their mathematical origin, EPs were introduced by Kato as isolated singularities of systems depending on one complex variable~\cite{Kato1995}, and they were recently classified into type I EPs and type II EPs~\cite{NHbook}. EPs of type I are degeneracies with or without algebraic singularities reminiscent of defective EPs and ONPs, while type II EPs are defined as points in the complex plane where the Jordan normal form is unstable, i.e., the eigenprojectors have a pole. While the appearance and existence of degeneracies reminiscent of type~I EPs have been studied extensively, type II EPs have hitherto been overlooked in the literature.

In this letter, we introduce a natural extension of type II EPs to higher dimensions, dubbed \emph{non-defective EPs}. We present that the correct criterion for detecting non-defective EPs is the form of the Hamiltonian matrix in the \emph{vicinity} of these degenerate points: Non-defective EPs are surrounded by defective EPs along certain intersections such that the Hamiltonian matrix casts a Jordan block in certain directions away from the non-defective EPs. Furthermore, we show that the Jordan decomposition is singular at these points when approaching the points from certain directions, emphasizing that non-defective EPs indeed are reminiscent of the type II EPs in Ref.~\onlinecite{NHbook}. To characterize the role of symmetries in witnessing NH degeneracies, we study the coexistence of defective and non-defective EP$n$s in two-, three- and four-band models in the presence of psH, ${\cal PT}$ or ${\cal CP}$ symmetry. Additionally, we show that symmetry-protected non-defective EP2s may also appear in models with non-Hermitian time-reversal symmetry. Finally, we find that defective EPs may coexist with ONPs instead of non-defective EPs when lifting symmetry constraints. We illustrate this finding with a fine-tuned example.

Finding possibilities to detect ONPs and (non-)defective EPs may pave the way to advance applications of Hermitian and non-Hermitian topological properties in various fields of research. For instance, topological lasers are one of the platforms which owe their success to either Hermitian~\cite{Wittek2017, Zeng2020} or non-Hermitian~\cite{Jean2017, Parto2018, Harari2018, Bandres2018} topological properties. 
Moreover, our systems, especially $\cal PT$-symmetric models, are experimentally feasible as they can be implemented in experimental optical setups with balanced gain and loss effectively~\cite{OzRoNoYa2019}. Exploring the role of EPs in $\cal PT$-symmetric optical systems has already unraveled many interesting phenomena, such as unidirectional invisibility~\cite{Regensburger2012}, induced mode-transition by encircling EPs~\cite{Doppler2016}, loss induced optical transparency~\cite{Guo2009}, and stable single-mode lasing in multi-mode optical setups~\cite{Liang2014, Hodaei2015}.

\paragraph*{Symmetry-stabilized (non-)defective EPs.}A generic $n$-band Hamiltonian can be decomposed as $\HH = d_{\mu} \Upsilon^{\mu}$,
where $\mu \in \{0, \cdots, n^{2}-1\}$, $d_{\mu}$ are continuously differentiable complex-valued functions of the lattice momentum $\bk$, $\Upsilon^{0}$ denotes the identity matrix of order $n$ and $\bolds{\Upsilon}$ is the basis of the $SU(n)$ group, which consists of three Pauli matrices when $n=2$, eight Gell-Mann matrices when $n=3$, and fifteen generalized Gell-Mann matrices when $n=4$ (see supplemental materials~(SM) for the explicit presentation of these matrices~\cite{SuppMat}). The Hamiltonian $\HH$ displays $\cal PT$ symmetry with generator ${\cal PT}$, $\cal CP$ symmetry with generator ${\cal CP}$, or $\rm psH$ with generator $\varsigma$, if it satisfies one of the following relations, namely,
\begin{align}
    {\cal PT}:& \quad
    \HH(\bk) = ({\cal PT}) \HH^{*}(\bk) ({\cal PT})^{-1} ,\\
    {\cal CP}:& \quad
    \HH(\bk) = -({\cal CP}) \HH^{*}(\bk) ({\cal CP})^{-1} ,\\
   {\rm psH}:& \quad
    \HH(\bk) = \varsigma \HH^{\dagger}(\bk) \varsigma^{-1}.
\end{align}
These symmetry considerations reduce the number of nonzero $d_{\mu}$ values. More precisely, for each basis matrix $\Upsilon^{\mu}$ merely a real or imaginary part of $d_{\mu}$ remains nonzero, i.e., only one real-valued function $d_{\mu}$ per each dimension of $\Upsilon^{\mu}$ survives~\cite{Sayyad2022}. Trivial band touching points occur when the traceless part of $\HH$ becomes a Null matrix~($[0]_{n \times n}$), i.e., all of the nonzero $d_{\mu}$ values for $\mu>0$ must vanish, which means that one needs to satisfy $n^2-1$ real constraints. For $n=2,3,4$, we have collected these $d_{\mu}$'s for each symmetry operation along side a choice for its generator in Tables~\ref{tab:symm_const}, \ref{tab:symm_const_3band}, and \ref{tab:symm_const_4band}, respectively.

The $n-1$ complex constraints to find defective EP$n$s can be expressed in terms of the traces and the determinant of $\HH$, which for two-, three-, and four-band models, respectively, read~\cite{Sayyad2022} 
\begin{align}
     n=2:&\quad \eta^{2b}=\tr[\HH]^{2}-4\det[\HH],\label{eq:const_2band}\\
     n=3:&\begin{cases}
     \eta^{3b} &= \frac{1}{2} \left( \tr[{\cal H}]^2-3 \tr[{\cal H}^2] \right) ,\\
\nu^{3b} &= \frac{1}{2} \left( 54 \det[{\cal H}]-5 \tr[{\cal H}]^3+9 \tr[{\cal H}] \tr[{\cal H}^2] \right),
     \end{cases}\label{eq:const_3band}
     \\
     n=4:&
     \begin{cases}
     \eta^{4b} &= -3 a c+b^2+12 d ,  \\
\nu^{4b} &= 27 a^2 d-9 a b c+2 b^3-72 b d+27 c^2 ,  \\
\kappa^{4b} &= a^3-4 a b+8 c,
     \end{cases}\label{eq:const_4band}
\end{align}
where $a=\tr[{\cal H}]$, $b=\big( (\tr[{\cal H}])^{2} - \tr[{\cal H}^2] \big) /2$, $d = \det[{\cal H}]$ and $
    c=\left(\tr[{\cal H}]^3-3 \tr[{\cal H}] \tr[{\cal H}^{2}]+2 \tr[{\cal H}^{3}]\right) /6$. We refer to the SM for details on how to derive these expressions \cite{SuppMat}.

In the presence of ${\cal PT}$, ${\cal CP}$ and psH symmetry, some of these constraints are automatically satisfied leaving us with exactly $n-1$ real constraints, cf. Tables~\ref{tab:symm_const}, \ref{tab:symm_const_3band} and \ref{tab:symm_const_4band}~\cite{Sayyad2022}.
It is notable that at trivial solutions (${\bf d} = 0$), all traces and the determinant of $\HH$ acquire zero values and subsequently, the constraints in Eqs.~(\ref{eq:const_2band})-(\ref{eq:const_4band}) are also satisfied. As a result, the trivial solutions mark non-defective EP$n$s with the binding signature that $\HH$ is diagonalizable at these points. Perturbing the system away from these non-defective points along intersections at which the constraints vanish brings the Hamiltonian into a non-diagonalizable structure. We set this behavior as a criterion to detect non-defective EPs. We note that this is opposed to the situation in which trivial solutions are isolated, and thus band touching points behave similarly to Hermitian degeneracies, i.e., ONPs.

The diagonalizability of the Hamiltonian at non-defective EPs enables us to map our NH Hamiltonians into their Hermitian counterparts with nodal points. In addition, having $n^2-1$ nonzero $d_{\mu}$'s as in Hermitian systems enforces non-defective EPs to always appear in pairs. This statement originates from the Poincar\'e-Hopf theorem~\cite{Mathai2017}, as the number of nonzero $d_\mu$ functions equals the dimension of the vector space~($n^{2}-1$), cf. last columns in Tables~\ref{tab:symm_const}, \ref{tab:symm_const_3band}, and \ref{tab:symm_const_4band}~\cite{Poincare}.

\begin{table}
     \centering
    \caption{Summarized symmetries, their generators and their associated constraints to find defective and non-defective EP2s in two-band systems.}
     \begin{tabular}{l|ll|ll|ll}
        \hline \hline 
        Symm.  &\qquad & Generator  &\qquad  & Constr. def. EP2s  &\qquad & Constr. non-def. EP2s\\
        \hline 
        \hline
              ${\cal PT}$
              & \qquad & ${\mathbbm 1}$
               & \qquad &  $\eta^{2b}_{R}=0$
        & \qquad & $d_{xR}=d_{yI}=d_{zR}=0$
       
         \\
              ${\cal CP}$
                 & \qquad & ${\mathbbm 1}$
                 & \qquad &  $\eta^{2b}_{R}=0$
        & \qquad & $d_{xI}=d_{yR}=d_{zI}=0$
        
         \\
           {\rm psH}
              & \qquad & ${\rm adiag}[1,1]$
                 & \qquad &  $\eta^{2b}_{R}=0$
        & \qquad & $d_{xR}=d_{yI}=d_{zI}=0$
     
         \\
          {TRS$^{\dagger}$}
            & \qquad & ${\rm adiag}[1,-1]$
             & \qquad & $\eta^{2b}_{R}=\eta^{2b}_{I}=0$ 
        & \qquad & $d_{xa}=d_{ya}=d_{za}=0$
       
          \\
        \hline 
     \end{tabular}
     \vspace{1ex}
     
     {\raggedright 
     Here we use either $d_{j}=d_{jR}+\i d_{jI}$ or $d_{j}=d_{js}+ d_{ja}$, where $d_{js}(d_{ja})$ is (anti-)symmetric with respect to $\bk \to -\bk$. $\eta^{2b}$ is given in Eq.~\eqref{eq:const_2band} with $\eta^{2b} = \eta^{2b}_R + \i \eta^{2b}_I$.
 \par} \label{tab:symm_const}
\end{table}

Aside from $\cal PT$, $\cal CP$ and $\rm psH$ symmetries, a particular non-Hermitian time-reversal symmetry, known as TRS$^{\dagger}$, in two-band systems may also give rise to realizing non-defective EPs. To evidently see this behavior, we recall that respecting TRS$^{\dagger}$ symmetry imposes $\HH(-\bk)= {\cal C}_{+} \HH^{\dagger}(\bk) {\cal C}^{\dagger}_{+}$.
This non-(momentum)-local transformation does not reduce the number of non-vanishing (real/imaginary) parts of $d_{\mu}$. However, when ${\cal C}_{+}=\i \sigma_{y}$, it enforces all symmetric parts of $d_{\mu}$ to become zero. We further know that right at time-reversal invariant momenta~(TRIM), e.g., $\bk_{\rm TRIM}=\{(0,0), (0,\pi),(\pi,0),(\pi,\pi) \}$ on the square Brillouin zone, anti-symmetric functions vanish. Therefore, at $\bk_{\rm TRIM}$ both real and imaginary parts of anti-symmetric $d_{\mu}$ functions become zero, which gives rise to observing non-defective EPs in the spectra of two-band TRS$^{\dagger}$-symmetric $\HH$, cf. Table~\ref{tab:symm_const}.

\begin{table*}
     \centering
   \caption{Summarized symmetries, their generators and their associated constraints to find defective and non-defective EP3s in three-band systems.}
     \begin{tabular}{l|ll|ll|ll}
        \hline \hline 
        Symm.  &\qquad & Generator  &\qquad  & Constr. def. EP3s &\qquad & Constr. non-def. EP3s \\
        \hline 
        \hline
              ${\cal PT}$
              & \qquad & ${\rm diag}[1,-1,1]$
               & \qquad &  $\eta^{3b}_{R}=0 \,\&\, \nu^{3b}_{R}=0$
        & \qquad & $d_{1R}=d_{4I}=d_{2I}=d_{5R}=d_{3R}=d_{6I}=d_{7R}=d_{8R}=0$
       
         \\
              ${\cal CP}$
                 & \qquad & ${\rm diag}[1,-1,1]$
                  & \qquad &  $\eta^{3b}_{R}=0 \,\&\, \nu^{3b}_{I}=0$
        & \qquad & $d_{1I}=d_{4R}=d_{2R}=d_{5I}=d_{3I}=d_{6R}=d_{7I}=d_{8I}=0$
       
         \\
           {\rm psH}
              & \qquad & ${\rm diag}[1,-1,1]$
              & \qquad & $\eta^{3b}_{R}=0 \,\&\, \nu^{3b}_{R}=0$
        & \qquad & $d_{1I}=d_{4I}=d_{2R}=d_{5R}=d_{3I}=d_{6I}=d_{7R}=d_{8R}=0$
        
          \\
        \hline 
     \end{tabular}
     \vspace{1ex}
     
     {\raggedright 
     Here we use $d_{j}=d_{jR}+\i d_{jI}$. Complex valued $\eta^{3b}$ and $\nu^{3b}$ constraints are given in Eq.~\eqref{eq:const_3band} with $\alpha^{3b} = \alpha^{3b}_R + \i \alpha^{3b}_I$ for $\alpha \in \{\eta, \nu\}$.
 \par} \label{tab:symm_const_3band}
\end{table*}

\begin{table*}
     \centering
    \caption{Summarized symmetries, their generators and their associated constraints to get defective and non-defective EP4s in four-band systems.}
     \begin{tabular}{l|ll|ll|ll}
        \hline \hline 
        Symm.  &\qquad & Generator  &\qquad  & Constr. def. EP4s &\qquad & Constr. non-def. EP4s \\
        \hline 
        \hline
              ${\cal PT}$
              & \qquad & ${\rm diag}[1,-1,1,-1]$
               & \qquad &  $\eta^{4b}_{R}=0 \,\&\, \nu^{4b}_{R}=0 \,\&\, \kappa^{4b}_{R}=0$
        & \qquad & \makecell{$d_{1R}=d_{2I}=d_{3R}=d_{4R}=d_{5I}=d_{6R}=d_{7I}=d_{8R}=0$\\$d_{9I}=d_{10I}=d_{11R}=d_{12I}=d_{13R}=d_{14R}=d_{15R}=0$}
       
         \\
         \hline
              ${\cal CP}$
                 & \qquad & ${\rm diag}[1,-1,1,-1]$
                  & \qquad &  $\eta^{4b}_{R}=0 \,\&\, \nu^{4b}_{R}=0 \,\&\, \kappa^{4b}_{I}=0$
        & \qquad & \makecell{$d_{1I}=d_{2R}=d_{3I}=d_{4I}=d_{5R}=d_{6I}=d_{7R}=d_{8I}=0$\\$
        d_{9R}=d_{10R}=d_{11I}=d_{12R}=d_{13I}=d_{14I}=d_{15I}=0$}
       
         \\
         \hline
           {\rm psH}
              & \qquad & ${\rm diag}[1,-1,1,-1]$
              & \qquad & $\eta^{4b}_{R}=0 \,\&\, \nu^{4b}_{R}=0 \,\&\, \kappa^{4b}_{R}=0$
        & \qquad & \makecell{$d_{1I}=d_{2R}=d_{3I}=d_{4I}=d_{5R}=d_{6I}=d_{7I}=d_{8R}=0$\\$d_{9I}=d_{10I}=d_{11R}=d_{12I}=d_{13R}=d_{14R}=d_{15R}=0$}
        
          \\
        \hline 
     \end{tabular}
     \vspace{1ex}
     
     {\raggedright 
     Here we use $d_{j}=d_{jR}+\i d_{jI}$. Complex valued $\eta^{4b}$, $\nu^{4b}$ and $\kappa^{4b}$ constraints are given in Eq.~\eqref{eq:const_4band} with $\alpha^{4b} = \alpha^{4b}_R + \i \alpha^{4b}_I$ for $\alpha \in \{\eta, \nu, \kappa\}$.
 \par} \label{tab:symm_const_4band}
\end{table*}

\begin{figure}[t!]
\centering

\includegraphics[width=0.99\columnwidth]{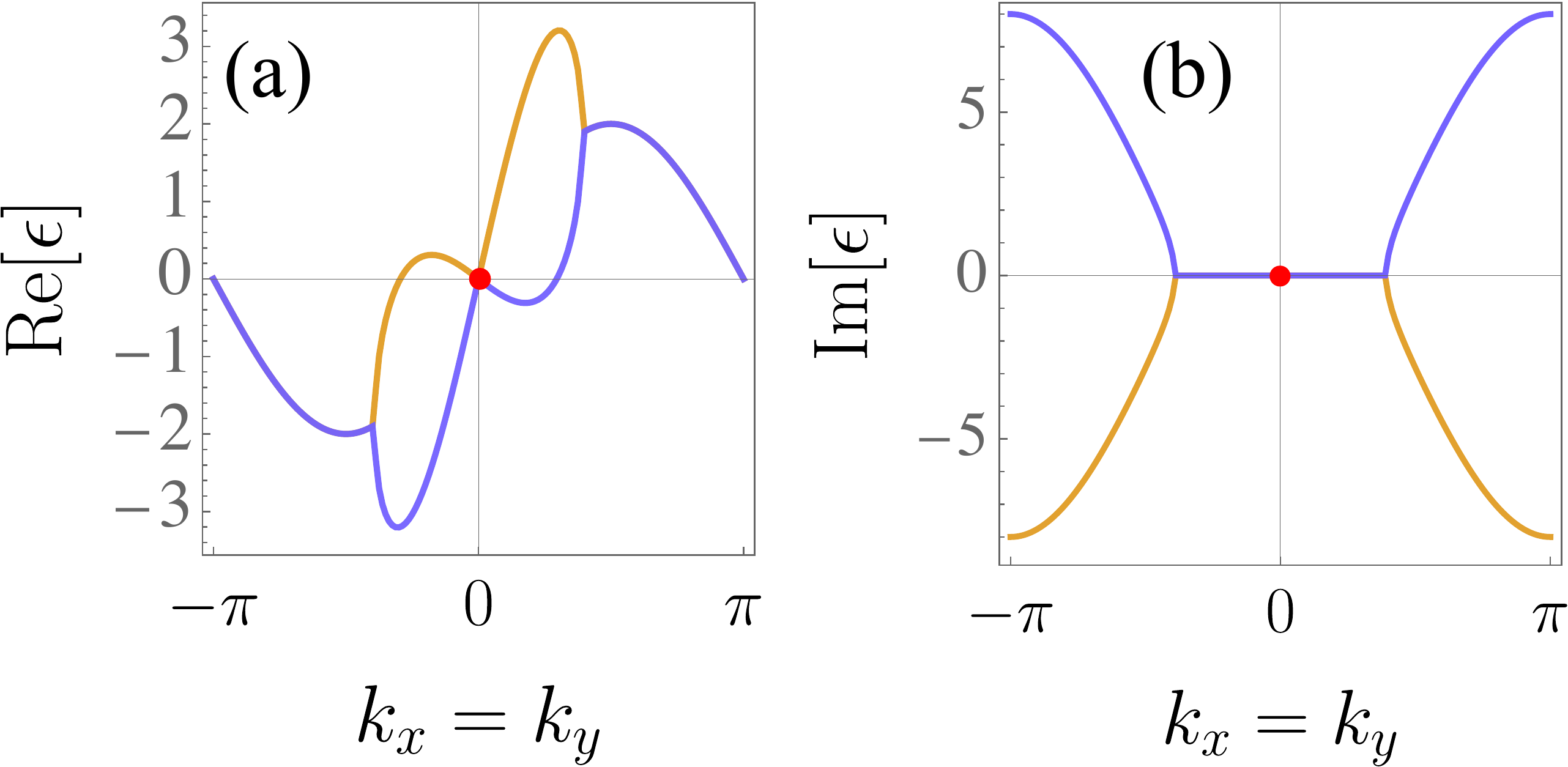}

\caption{\label{fig:spectra2b} Real part~(a) and imaginary part~(b) of spectra for Eq.~\eqref{eq:H2bpt} along $k_{x}=k_{y}$ with $k_{z}=\pi/2$. Red points indicate non-defective EPs. Here we set $t=V=\lambda_{0}=1.0$.}
\end{figure}

Before moving on to examples, we note that the different number of constraints that need to be satisfied to find symmetry-protected defective and non-defective EP$n$s also result in a different codimension of these EPs. Here, the codimension is given by the difference between the total dimension of the system and the dimension of the exceptional feature. Equivalently, the codimension corresponds precisely to the number of nonvanishing constraints. In particular, while the presence of ${\cal PT}$, ${\cal CP}$ and psH symmetries reduces the number of real constraints for finding defective EP$n$s to $n-1$, the number of real constraints to detect non-defective EP$n$s is $n^2-1$. As a consequence, in the case of $n=2$, defective EP2s have codimension \emph{one}, whereas non-defective EP2s have codimension \emph{three}. Therefore, the latter appear as points in three-dimensional systems, whereas defective EP2s appear as two-dimensional surfaces. For the TRS$^{\dagger}$ invariant two-band model, the codimension is \emph{two}, and hence the defective EP2s are curves connected at the TRIMs.

\paragraph*{Examples for the coexistence of defective and non-defective EPs.}We start with introducing a two-band ${\cal PT}$-symmetric Weyl-like tight-binding model described by
\begin{align}
    \HH_{\cal PT}^{2b} &= d_{0}\Upsilon^0 + d_{xR}\Upsilon^1 + \i d_{yI} \Upsilon^2 + d_{zR} \Upsilon^3 \nonumber \\
    &=
    2 \lambda_0 \sin{\left(k_x\right)} \Upsilon^0+2t\sin{\left(k_x\right)}\Upsilon^1 + 2t\sin{\left(k_y\right)}\Upsilon^3 \nonumber
    \\
    & + \i\left\{ 2t\cos{\left(k_z\right)} + 2V\left[2-\cos{\left(k_x\right)}-\cos{\left(k_y\right)}\right]\right\} \Upsilon^2.
    \label{eq:H2bpt}
\end{align}
Here $\lambda_{0}$, $t$ and $V$ are real-valued parameters.
The real and imaginary parts of the band structure are shown in Figs.~\ref{fig:spectra2b}(a) and (b), respectively.

\begin{figure}[t!]
\centering
\includegraphics[width=0.5\columnwidth]{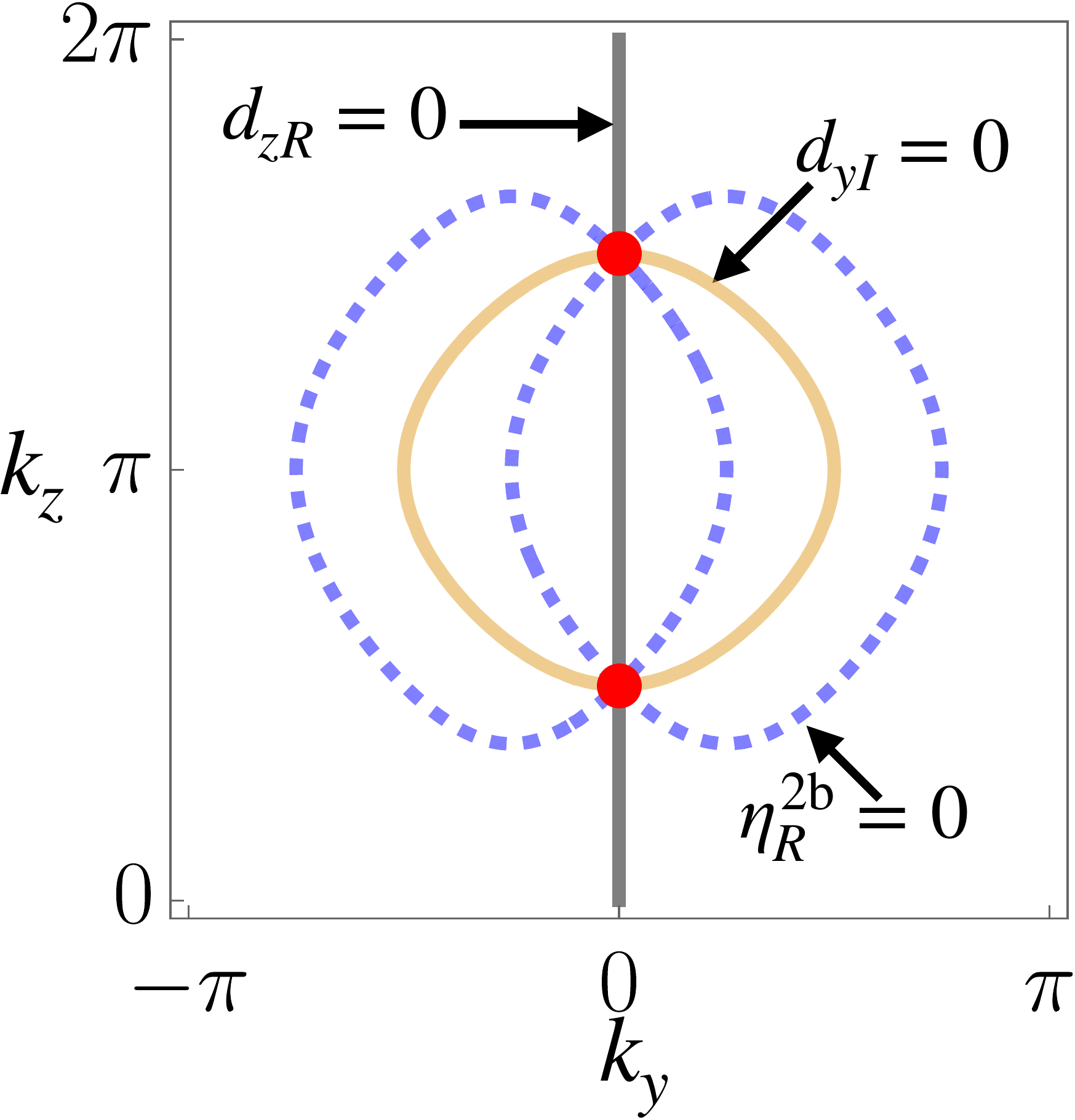}
\caption{\label{fig:const2b} 
Solutions of $d_{yI}$~(orange), $d_{zR}$~(gray), and $\eta_{R}^{2b}$~(dashed blue) at $k_{x}=0$, which is a solution to $d_{xR} = 0$. Red points at $\bk=(0,0,\pm \pi/2)$ indicate the intersection between solutions to $\eta_{R}^{2b}=d_{yI}=d_{zR}=0$, and are non-defective EPs.  Here we set $t=V=\lambda_{0}=1.0$.
}
\end{figure}

Non-defective EPs appear when all components of the Hamiltonian, except $d_{0}$, vanish. More specifically, these degeneracies emerge when the solutions of $d_{xR}=0$~(at $k_{x}=n \pi$ with $n \in \mathbb{Z}$), $d_{yI}=0$~(orange curves in Fig.~\ref{fig:const2b}), and $d_{zR}=0$~(grey line) intersect. Red points at $\bk
=\left(0,0,\pm \pi /2 \right)$ in Fig.~\ref{fig:const2b} exemplify such solutions. 
Note that the criterion for detecting non-defective EPs is satisfied for the red points in Fig.~\ref{fig:const2b} as they are surrounded by defective EPs~(dashed blue curves), residing on $\eta^{2b}_{R}
=0$, where $\eta^{2b}_{R}$ is the real part of $\eta^{2b}$ in Eq.~\eqref{eq:const_2band}.

Along the defective EPs, defined by $d_{xR}^2+d_{zR}^2=d_{yI}^2$, the Jordan decomposition reads $\HH_{\PT}^{2b} = SJ S^{-1}$ with $S=(s_1, s_2)$, where
\begin{align}
    s_1 &= \begin{pmatrix} \frac{\sqrt{t^4\left[2-\cos{\left(2k_x\right)}-\cos{\left(2k_y\right)}\right]}-\sqrt{2} t^2\sin{\left(k_x\right)}}{\sqrt{2} t^2\sin{\left(k_y\right)}} \\1  \end{pmatrix},
    \\
    s_2 &= \begin{pmatrix} \frac{\sqrt{2}\sqrt{t^4\left[2-\cos{\left(2k_x\right)}-\cos{\left(2k_y\right)}\right]}-2 t^2\sin{\left(k_x\right)}}{4 t^3\sin{\left(k_y\right)}^2} \\0\end{pmatrix},
    \\
    J &= \begin{pmatrix} 2\lambda_0 \sin{\left(k_x\right)}&1 \\0&2 \lambda_0 \sin{\left(k_x\right)}\end{pmatrix}.
\end{align}
The second column in the transformation matrix, $s_2$, has a singularity when $k_x$ and $k_y$ goes to $n\pi$ simultaneously, corresponding exactly to the non-defective EP. Hence, the Jordan decomposition is unstable at the non-defective EPs. 

We further note that the defective EPs separate two regions in the real part of spectrum, where $\text{Re}[\Delta \epsilon] = 0$ and $\text{Re}[\Delta \epsilon] \neq 0$ with $\Delta \epsilon$ the difference between the two energy bands as shown in Fig.~\ref{fig:spectra2b}(a). Regions where $\text{Re}[\Delta \epsilon] = 0$ are sometimes referred to as NH bulk real-Fermi states, which merely appear in NH systems~\cite{NHreview}. In the SM, we show that besides these bulk Fermi states, this model also hosts states on the boundary~\cite{SuppMat}. Therefore, there is a coexistence between defective and non-defective EPs as well as between bulk Fermi states and boundary states.

\begin{figure}[t!]
\centering
\includegraphics[width=0.99\columnwidth]{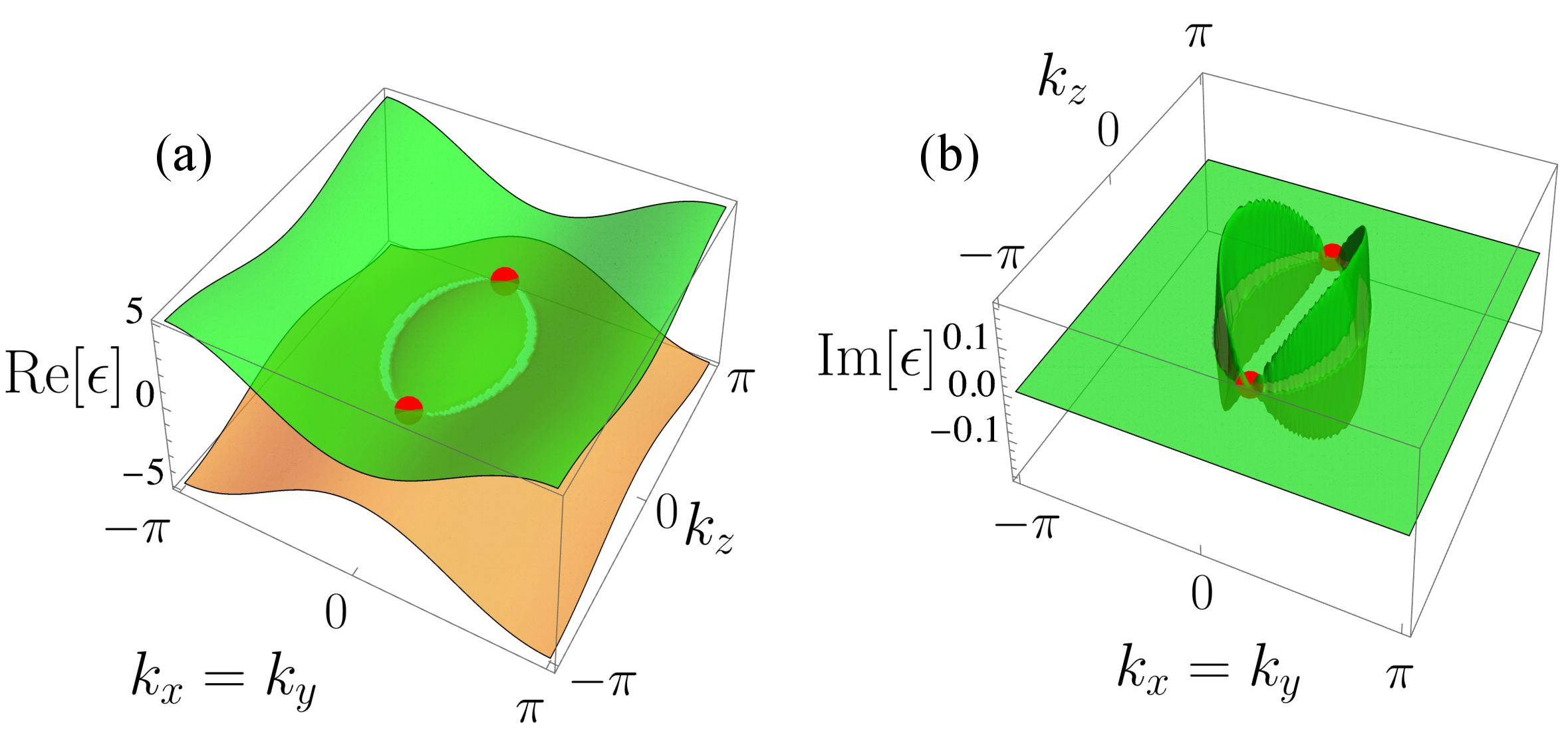}
\caption{\label{fig:4bdirac} 
Real~(a) and imaginary~(b) parts of band structure for $ \HH_{\rm psH}^{4b}$ in Eq.~\eqref{eq:H4band_dirac} along $k_{x}=k_{y}$ and $k_{z}$. Note that both bands are doubly degenerate. Red points indicate non-defective EPs at $\bk=(0,0,\pm k_{0})$. Here we set $t=t_{z}=1.0$, $\lambda_{Ixx}=\lambda_{Ixy}=0.15$, $m'_{I}=-0.27$ and $k_{0}=\pi/2$.
}
\end{figure}

Let us now turn to a four-band model. We consider a Dirac-like $\rm psH$-symmetric model described by
\begin{align}
    \HH_{\rm psH}^{4b}= &
   t \left[\cos{\left(k_{x}\right)} + \cos{\left(k_y\right)}-2\right]   \left(\frac{2}{\sqrt{3}} \Upsilon^{14} + \sqrt{\frac{2}{3}} \Upsilon^{15}\right)
        \nonumber\\
    &
  +t_{z}\left[\cos{\left(k_{z}\right)}-\cos{\left( k_{0}\right)} \right]\left[\frac{2}{\sqrt{3}} \Upsilon^{14} + \sqrt{\frac{2}{3}} \Upsilon^{15}\right]
        \nonumber\\
    &
  +  \i \lambda_{I xy} \sin{\left(k_{y}\right)} (\Upsilon^{3} + \Upsilon^{4})
     \nonumber\\
     &
 +   \i \lambda_{I xx} \sin{\left(k_{x}\right)} (\Upsilon^{9} + \Upsilon^{10})
        \nonumber\\
    &
 +   \i m'_{I} \sin{\left(k_{z}\right)} \left[\cos{\left(k_{x}\right)} - \cos{\left(k_{y}\right)}\right] \left(\Upsilon^{7} - \Upsilon^{12}\right)
 \label{eq:H4band_dirac}
 ,
\end{align}
where $t$, $t_{z}$, $\lambda_{I xy}$, $\lambda_{I xx}$, and $m_{I}^{'}$ are real-valued parameters.
This model is a $\rm psH$ generalization of the tight-binding model studied in Ref.~\onlinecite{Kargarian2016}. 
The trivial band touching points for obtaining the null form of the traceless part of $\HH_{\rm psH}^{4b}$ in Eq.~\eqref{eq:H4band_dirac} are located at
$\bk =\bk_{\rm psH}
=(0,0,\pm k_{0})$. Right at these points and on lines connecting these points, constraints for realizing EP4s, summarized in Table~\ref{tab:symm_const_4band} and given in Eq.~\eqref{eq:const_4band}, are also satisfied.
Hence, points at $\bk=\bk_{\rm psH}$ in our $\rm psH$-symmetric model are non-defective EP4s. We present these nodal points~(red spheres) in the band structure of $\HH_{\rm psH}^{4b}$ in Fig.~\ref{fig:4bdirac}. The real part of the spectra~(a) shows that two
arc-shaped surfaces with (non)zero real (imaginary) parts are terminated by the non-defective EPs as well as the defective exceptional lines. These surfaces are the aforementioned NH bulk real-Fermi
surfaces. While our model hosts bulk Fermi surfaces, boundary states are not stable, similar to its Hermitian counterpart~\cite{Kargarian2016}. 

\begin{figure}[b!]
\centering
\includegraphics[width=0.99\columnwidth]{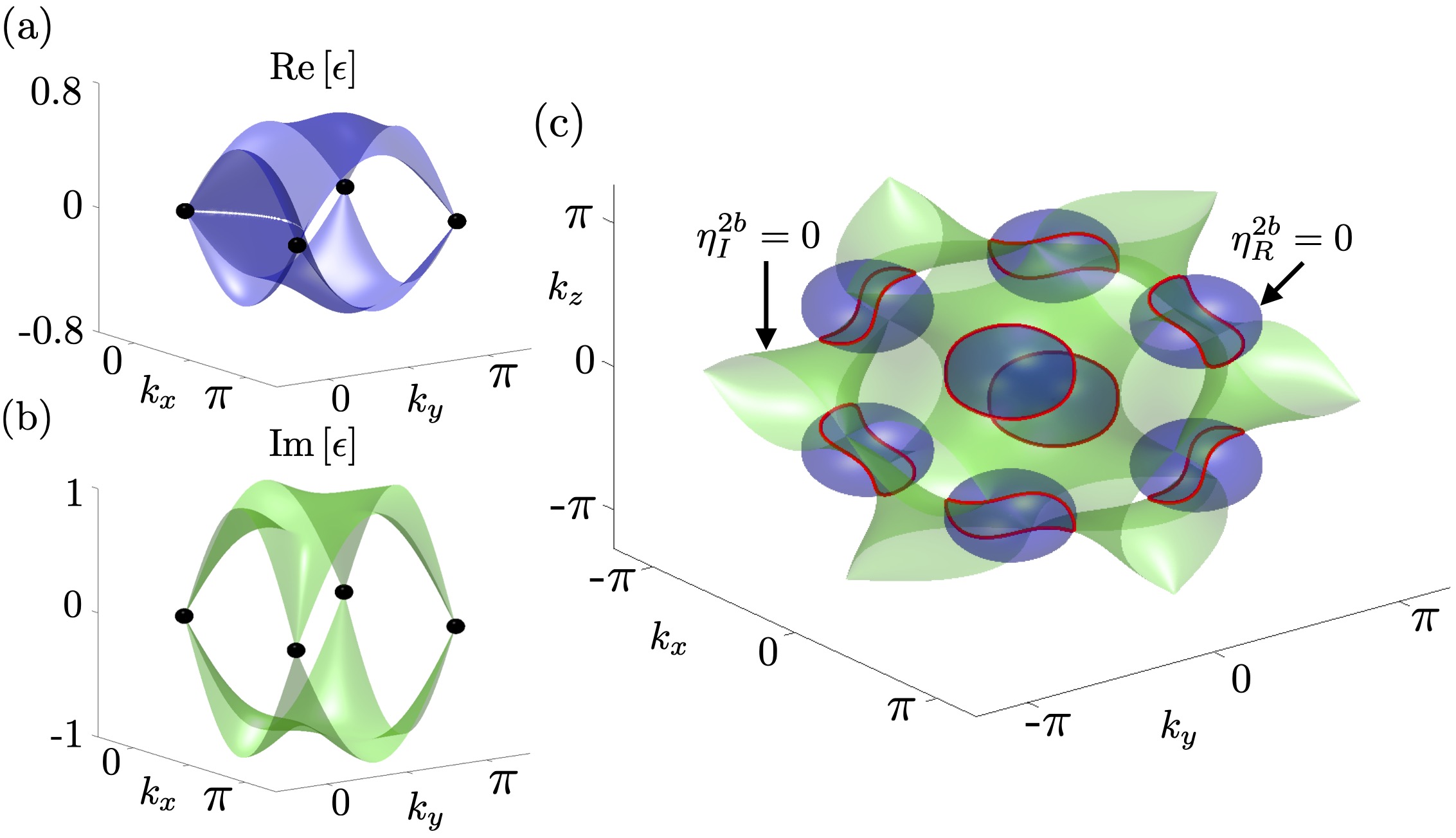}
\caption{Real~(a) and imaginary~(b) parts of band structure for $ \HH_{\rm ONP}^{\rm 2b}$ in Eq.~\eqref{eq:H2bOP} along $k_{z}=\pi$. (c) Solutions of the real~(blue spheres) and imaginary~(green manifold) parts of $\eta^{2b}$ in Eq.~\eqref{eq:const_2band}.
 Red closed lines in (c) present the intersection of these manifolds and correspond to defective EPs. Black points in panels (a), (b) indicate ONPs. For visibility purposes, merely a corner of Brillouin zone is shown in panels (a), (b).
\label{fig:2bOP}
}
\end{figure}

\paragraph*{Searching for the coexistence of ONPs and defective EPs.}So far, we have explored how the presence of one of the ${\cal PT}$, ${\cal CP}$ or $\rm psH$ symmetries leads to the general coexistence of defective and non-defective EPs.
In the remaining, we address whether ONPs may also exist in NH spectra. 
For this purpose, we lift (discrete) symmetry restrictions such that some (or all) $d_{\mu}$'s have real and imaginary parts. We emphasize that in this situation, in contrast to the EPs in the previous sections, band degeneracies are generally unstable to small perturbations \footnote{More precisely, twofold degeneracies are protected by composite symmetries consisting of multiple symmetry operations~\cite{Sayyad2022d}. Respecting all these symmetry operations might be easily violated upon introducing perturbations.}. This is because finding solutions to ${\bf d}=0$ generally requires satisfying $n^2-1$ complex constraints. Consequently, the appearance of ONPs in NH models is vulnerable to the fine-tuning of parameters. Nevertheless, we show in the following that this setting provides a platform to observe ONPs. 

To illustrate our idea, we introduce a two-band model given by
\begin{align}
    \HH_{\rm ONP}^{2b} &= \sin (k_x) \left[ 1/2+\i \cos (k_y) \right] \Upsilon^1 \nonumber\\
    &+ \sin( k_y) \left[ 1/2+\i\cos (k_z) \right] \Upsilon^2 \nonumber\\
    &+ \sin (k_z) \left[ 1/2+\i\cos (k_x) \right] \Upsilon^3
    .
    \label{eq:H2bOP}
\end{align}
Fig.~\ref{fig:2bOP} exhibits real~(a) and imaginary~(b) parts of the energy dispersion of this system along $k_{z}=\pi$. The black points in these panels mark real and imaginary band touching points, located at $k_{x,y,z}^{\rm ONP}=\pm n\pi$. Even though at these momenta the traceless part of $\HH_{\rm ONP}^{2b}$ becomes a null matrix, we emphasize that ${\bf k}={\bf k}^{\rm ONP}$ indicates the location of ONPs and not non-defective EPs. The reason for this statement lies in the fact that the criterion for the emergence of non-defective EPs is not satisfied, i.e., no defective EPs reside close to ${\bf k}^{\rm ONP}$. This can be seen from Fig.~\ref{fig:2bOP}~(c) in which we present defective EPs, red curves, as the intersection between $\text{Re}[\eta^{2b}] = 0$, blue spheres, and $\text{Im}[\eta^{2b}]=0$, green manifold. Fig.~\ref{fig:2bOP}~(c) reveals that defective EPs do not cross ${\bf k}={\bf k}^{\rm ONP}$ and thus the black points at ${\bf k}^{\rm ONP}$ indeed correspond to ONPs. Consequently, $H^{2b}_{\text{ONP}}$ is diagonalizable at ${\bf k}={\bf k}^{\rm ONP}$ and in any small neighborhood surrounding the point.

Aside from these ONPs in the momentum-space, introducing models that host boundary states connecting ONPs in NH systems is theoretically feasible. We present an example of such a model in the SM~\cite{SuppMat}.

\paragraph*{Conclusion.}Despite the intense focus on NH systems in recent studies, the possibility of realizing different types of EPs has hitherto been overlooked. The present work shows that two different types of EPs, dubbed defective and non-defective EPs, may coexist in various setups of physical importance. Concretely, we show that non-defective EPs are stabilized by certain symmetries, including $\cal PT$, $\cal CP$, $\rm psH$, and time-reversal symmetry. To resolve the confusion in the current literature, where non-defective EPs are mixed up with ONPs, we have in this work introduced a clear criterion to distinguish these concepts. We also highlight this difference in example models.

Even though the models we discuss in this work are experimentally tractable, we do not expect a difference in the experimental signatures of non-defective EPs and ONPs when single-particle, non-interacting Hamiltonians are explored. Investigating interacting/coupled systems may allow identifying distinct footprints of non-defective EPs and ONPs in experiments.

\paragraph*{Acknowledgements.}M.S. is supported by the Swedish Research Council~(VR) and the Wallenberg Academy Fellows program of the Knut and Alice Wallenberg Foundation. L.R. acknowledges the support from the Knut and Alice Wallenberg foundation under grant no. 2017.0157.

\bibliography{Coex_ONP_EP.bib}

\appendix

\renewcommand{\thefigure}{S\arabic{figure}}
\setcounter{figure}{0} 

\section{Bases matrices for two-, three-, and four-band systems }\label{app:bases}

\subsection{Basis matrices for two-band systems} 

The basis matrices for two-band systems are Pauli matrices which read
\begin{align}
\Upsilon^{1} =
\begin{pmatrix}
0 & 1 \\
1 & 0
\end{pmatrix}
,\, \,
\Upsilon^{2} =
\begin{pmatrix}
0 & -\i \\
\i & 0
\end{pmatrix}
,\, \,
\Upsilon^{3} =
\begin{pmatrix}
1 & 0 \\
0 & -1
\end{pmatrix}.
\end{align}

\subsection{Basis matrices for three-band systems} \label{app:gellmann_su3}

The basis matrices for three-band systems are the Gell-Mann matrices, that span the Lie algebra of the SU(3) group,
\begin{align}
\Upsilon^{1}&=
 \begin{pmatrix}
0  & -\i & 0 \\
\i & 0 & 0 \\
0 & 0 & 0
\end{pmatrix}, \quad 
\Upsilon^{2}=
 \begin{pmatrix}
0  & 0& -\i  \\
0& 0 & 0 \\
\i & 0 & 0
\end{pmatrix} ,\\ 
\Upsilon^{3}&= 
\begin{pmatrix}
0  & 0 & 0 \\
0 & 0 & -\i \\
0 & \i & 0
\end{pmatrix}, \quad 
\Upsilon^{4}= 
\begin{pmatrix}
0  & 1 & 0 \\
1 & 0 & 0 \\
0 & 0 & 0
\end{pmatrix}, \\
\Upsilon^{5}&= 
 \begin{pmatrix}
0  & 0 & 1 \\
0 & 0 & 0 \\
1 & 0 & 0
\end{pmatrix}, \quad 
\Upsilon^{6}=
 \begin{pmatrix}
0  & 0 & 0 \\
0 & 0 & 1 \\
0 & 1 & 0
\end{pmatrix},\\
\Upsilon^{7}&=
  \begin{pmatrix}
1 & 0 & 0 \\
0 & -1 & 0  \\
0 & 0 & 0
\end{pmatrix}, \quad 
\Upsilon^{8}= 
\begin{pmatrix}
\frac{1}{\sqrt{3}}  & 0 & 0 \\
0 & \frac{1}{\sqrt{3}} & 0 \\
0 & 0 & -\frac{2}{\sqrt{3}}
\end{pmatrix}.
\end{align}

\subsection{Basis matrices for four-band systems} \label{app:gellmann_su4}

The basis matrices for four-band systems are the generalized Gell-Mann matrices, that span the Lie algebra of the SU(4) group,
\begin{align}
\Upsilon^{1}&= 
\begin{pmatrix}
0 & -\i & 0 & 0 \\
\i & 0 & 0 & 0 \\
0 & 0 & 0 & 0 \\
0 & 0 & 0 & 0 \\
\end{pmatrix}
,\quad 
\Upsilon^{2}= 
\begin{pmatrix}
0 & 0 &-\i  & 0 \\
0 & 0 & 0 & 0 \\
\i & 0 & 0 & 0 \\
0 & 0 & 0 & 0 \\
\end{pmatrix}
,\\
\Upsilon^{3}&= 
\begin{pmatrix}
0 & 0 & 0 &-\i   \\
0 & 0 & 0 & 0 \\
0 & 0 & 0 & 0 \\
\i & 0 & 0 & 0 \\
\end{pmatrix}
,\quad 
\Upsilon^{4}= 
\begin{pmatrix}
0 & 0 & 0 &0   \\
0 & 0 & -\i & 0 \\
0 & \i & 0 & 0 \\
0 & 0 & 0 & 0 \\
\end{pmatrix}
,\\
\Upsilon^{5}&= 
\begin{pmatrix}
0 & 0 & 0 &0   \\
0 & 0 & 0 & -\i \\
0 & 0 & 0 & 0 \\
0 & \i & 0 & 0 \\
\end{pmatrix}
,\quad 
\Upsilon^{6}= 
\begin{pmatrix}
0 & 0 & 0 &0   \\
0 & 0 & 0 & 0 \\
0 & 0 & 0 & -\i \\
0 & 0 & \i & 0 \\
\end{pmatrix}
,\\ 
\Upsilon^{7}&= 
\begin{pmatrix}
0 & 1 & 0 &0   \\
1 & 0 & 0 & 0 \\
0 & 0 & 0 & 0 \\
0 & 0 & 0 & 0 \\
\end{pmatrix}
,\quad 
\Upsilon^{8}= 
\begin{pmatrix}
0 & 0 & 1 &0   \\
0 & 0 & 0 & 0 \\
1 & 0 & 0 & 0 \\
0 & 0 & 0 & 0 \\
\end{pmatrix}
,\\
\Upsilon^{9}&= 
\begin{pmatrix}
0 & 0 & 0 &1   \\
0 & 0 & 0 & 0 \\
0 & 0 & 0 & 0 \\
1 & 0 & 0 & 0 \\
\end{pmatrix}
,\quad
\Upsilon^{10}= 
\begin{pmatrix}
0 & 0 & 0 & 0   \\
0 & 0 & 1 & 0 \\
0 & 1 & 0 & 0 \\
0 & 0 & 0 & 0 \\
\end{pmatrix}
,\\
\Upsilon^{11}&= 
\begin{pmatrix}
0 & 0 & 0 & 0   \\
0 & 0 & 0 & 1 \\
0 & 0 & 0 & 0 \\
0 & 1 & 0 & 0 \\
\end{pmatrix}
,\quad
\Upsilon^{12}= 
\begin{pmatrix}
0 & 0 & 0 & 0   \\
0 & 0 & 0 & 0 \\
0 & 0 & 0 & 1 \\
0 & 0 & 1 & 0 \\
\end{pmatrix}
,\\
\Upsilon^{13}&= 
\begin{pmatrix}
1 & 0 & 0 & 0   \\
0 & -1 & 0 & 0 \\
0 & 0 & 0 & 0 \\
0 & 0 & 0 & 0 \\
\end{pmatrix}
,\,
\Upsilon^{14}= 
\begin{pmatrix}
\frac{1}{\sqrt{3}} & 0 & 0 & 0   \\
0 & \frac{1}{\sqrt{3}} & 0 & 0 \\
0 & 0 & -\frac{2}{\sqrt{3}} & 0 \\
0 & 0 & 0 & 0 \\
\end{pmatrix}
,\\
\Upsilon^{15}&= 
\begin{pmatrix}
\frac{1}{\sqrt{6}} & 0 & 0 & 0   \\
0 & \frac{1}{\sqrt{6}} & 0 & 0 \\
0 & 0 & \frac{1}{\sqrt{6}} & 0 \\
0 & 0 & 0 & -\sqrt{\frac{3}{2}} \\
\end{pmatrix}.
\end{align}

\begin{figure}[t]
\centering

\includegraphics[width=0.99\columnwidth]{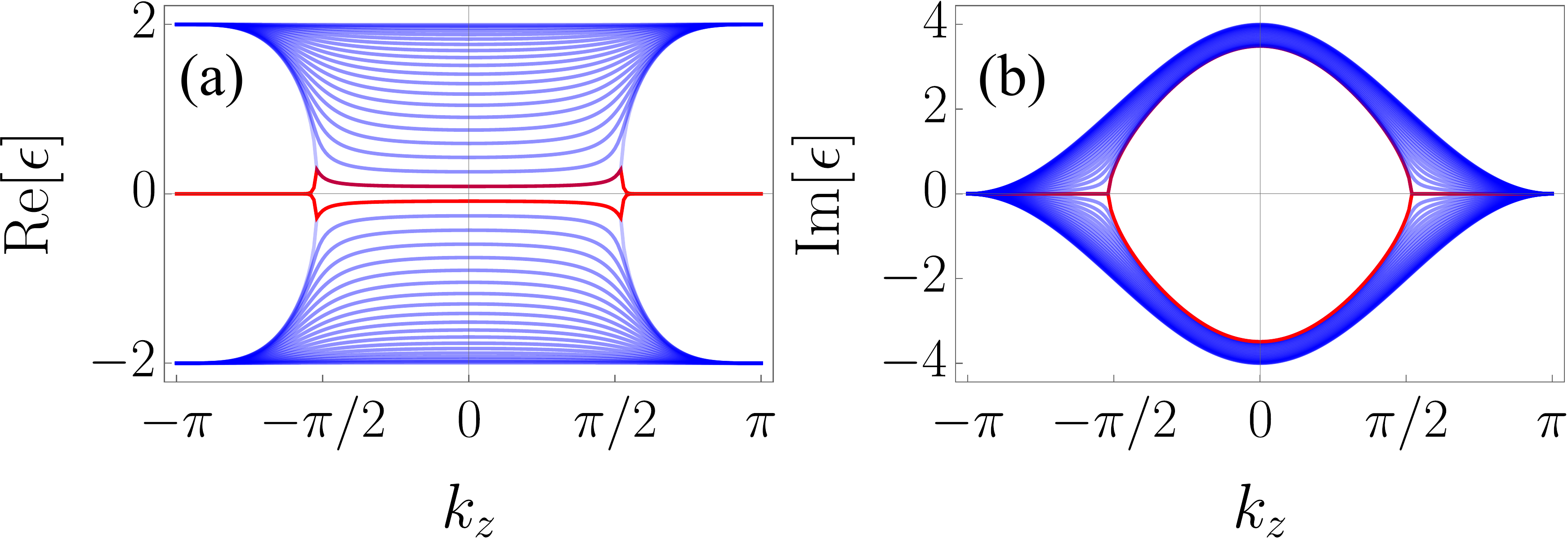}

\caption{\label{fig:spectra2b_rspace} Real parts~(a) and imaginary parts~(b) of spectra for Eq.~\eqref{eq:H2bpt} with open boundary condition in the $y$ direction and $k_{x}=0$. Red lines present boundary states. Here we set $t=V=\lambda_{0}=1.0$.}
\end{figure}

\section{Derivation of constraints to find EP$n$s}

Here we briefly summarize how to derive Eqs.~(4)-(6) in the main text, where these constraints were originally derived in Ref.~\onlinecite{Sayyad2022}. There it was shown that the characteristic polynomial of an $n$-band NH matrix can be expressed in terms of its determinant and traces. Indeed, for two-, three- and four-band matrices, these polynomials read
\begin{align}
     F^{2b}(\lambda) & = \lambda^2 - \tr[\mathcal{H}]\lambda + \det[\mathcal{H}]=0, \\
     F^{3b}(\lambda) &=\lambda^3 - \tr[\mathcal{H}]\lambda^2 + \frac{\tr[\mathcal{H}]^2 - \tr[\mathcal{H}^2]}{2} \lambda - \det[\mathcal{H}] = 0, \\
     F^{4b}(\lambda) &=\lambda^4 - a \lambda^3 + b \lambda^2 - c \lambda + d = 0,
\end{align}
where
\begin{align}
    a & = \tr[\mathcal{H}], \quad b = \frac{\tr[\mathcal{H}]^2 - \tr[\mathcal{H}^2]}{2}, \quad d = \det[\mathcal{H}], \\
    c & = \frac{\tr[\mathcal{H}]^3- 3 \tr[\mathcal{H}] \tr[\mathcal{H}^2] + 2 \tr[\mathcal{H}^3]}{6},
\end{align}
and $\lambda$ are the eigenvalues.

To find degeneracies, the discriminant $D[\mathcal{H}]$ of these characteristic polynomials need to be set to zero. The discriminants read
\begin{align}
    D[\mathcal{H}^{2b}] &= \tr[\mathcal{H}]^2 - 4 \det[\mathcal{H}], \label{eq:disc_two} \\
    D[\mathcal{H}^{3b}] &= - \frac{1}{27} [4 (\eta^{3b})^3 + (\nu^{3b})^2], \label{eq:disc_three}\\
    D[\mathcal{H}^{4b}] &=\frac{1}{27} [4 (\eta^{4b})^3 - (\nu^{4b})^2],
\end{align}
where $\eta^{3b}$ and $\nu^{3b}$, and $\eta^{4b}$ and $\nu^{4b}$ are given in Eqs.~(5) and (6) in the main text, respectively. From here, we immediately see that setting the discriminants in Eqs.~\eqref{eq:disc_two} and \eqref{eq:disc_three} to zero gives us the constraints in Eqs.~(4) and (5) in the main text, respectively. In the case of the four-band model, we note that for all roots of the discriminant to coincide, not only $\eta^{4b} = 0$ and $\nu^{4b} = 0$ need to be satisfied, but also $\kappa^{4b} = 0$, where $\kappa^{4b}$ is defined in Eq.~(6) in the main text. We refer to Ref.~\onlinecite{Sayyad2022} for a more detailed discussion on this point.

\section{Spectra of the two-band model with open boundary condition}
In addition to the properties of momentum-dependent spectra for ${\cal H}_{\cal PT}^{\rm 2b}$ in Fig.~\ref{fig:spectra2b}, we present real~(a) and imaginary~(b) parts of the energy dispersion with open boundary condition in the $y$ direction plotted in Fig.~\ref{fig:spectra2b_rspace}. The figure exhibits boundary states, red lines, well-separated from the bulk states~(blue lines) when $|k_{z}| > \pi/2$. 

\begin{figure}[t!]
\centering

\includegraphics[width=0.99\columnwidth]{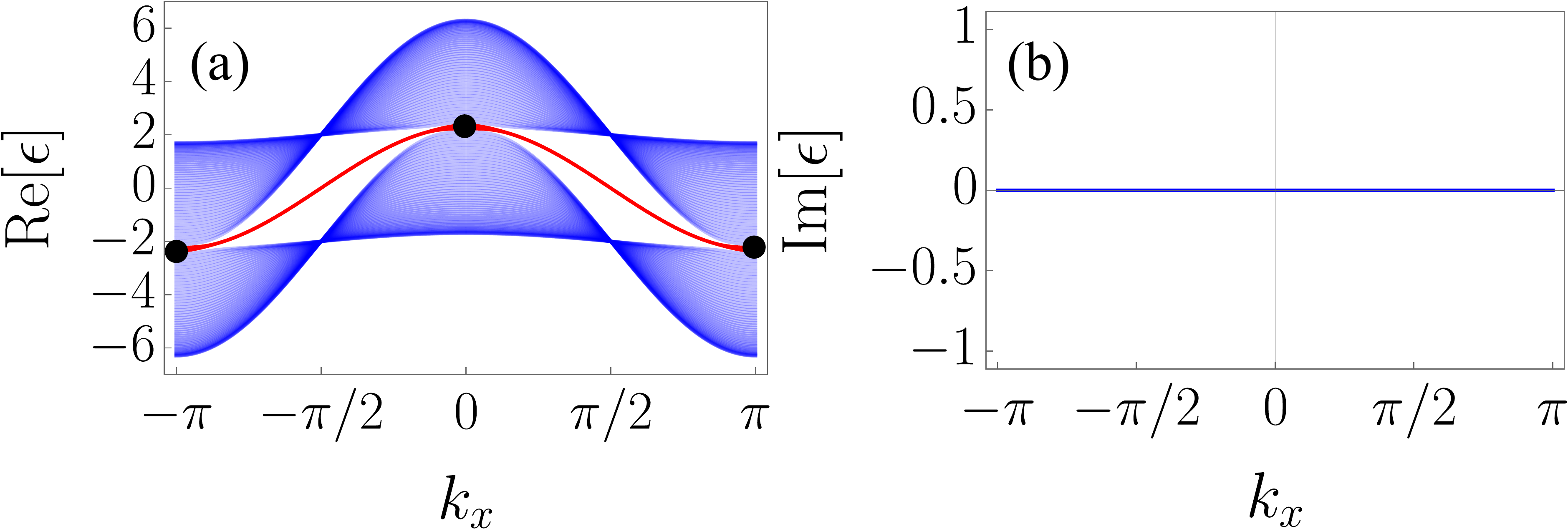}

\caption{\label{fig:spectra2b_mymodel} Real parts~(a) and imaginary parts~(b) of spectra for Eq.~\eqref{eq:Hedge2b} with open boundary condition in the $y$ direction and at $k_{z}=0$. Red lines present boundary states. Black points mark ONPs. Here we set $t=V=1.0, \lambda_{0}=2.3$.}
\end{figure}

\section{Realizing Hermitian boundary states in non-Hermitian systems}
Here we present a non-Hermitian tight-binding model hosting Hermitian boundary states, which connects ONPs with zero imaginary parts.

Our model Hamiltonian reads
\begin{align}
    \HH_{\rm edg}^{\rm 2b} =& 
    \lambda_{0}  \cos(k_{x})\Upsilon^{0}
    -\i V \left[  1- \cos(k_{z})  \right]\Upsilon^{1}
    \nonumber \\
    &
    +\left[ 2V \cos(k_y) -2t  \cos(k_{x})  \right]\Upsilon^{1}
        \nonumber \\
    &
    -2 t \sin(k_y) \Upsilon^{2} -2t \sin(k_{z}) \Upsilon^{3},
    \label{eq:Hedge2b}
\end{align}
where $\lambda_{0}$, $t$ and $V$ are real-valued coupling constants. Along $k_{z}=0$, the above Hamiltonian is fully Hermitian. As a result, nodal points, which live on the $(k_x,k_y)$ plane, are band touching points with zero imaginary parts. For instance, these ordinary nodal points appear at $\bk_{\rm ONPs}=(\pm n \pi, \pm n \pi, 0)$ with $n \in \Z$ when $t=V$. Considering the open boundary condition along the $y$ axis and at $k_{z}=0$ results in mid-gap boundary states, red lines in Fig.~\ref{fig:spectra2b_mymodel}(a), which connects $\bk_{\rm ONPs}$.

\end{document}